# Engineered Inclined Energy Landscapes Enabling Free Flow of Magnetic Microstructures for Artificial Neuron Applications

Anmol Sharma[a], Ranjeet Kumar Brajpuriya[a], Vivek K. Malik[b], Vishakha Kaushik[c], Sachin Pathak[a]*

[a]Department of Physics, UPES Dehradun, Uttarakhand, India, 248007
[b]Department of Physics, Indian Institute of Technology Roorkee, Roorkee 247667, India
[c]Materials & Nano Engineering Research Laboratory, Department of Physics, School of Physical Sciences, DIT University, Dehradun- 248009, India

*Corresponding author: s.pathak@ddn.upes.ac.in (Sachin Pathak)

**Abstract:** Spintronic-based brain-inspired neuromorphic computing has recently attracted significant attention due to the exceptional properties of magnetic microstructures, including nanoscale dimensions, high stability, and low energy consumption. Despite these advantages, the practical integration of such microstructures into functional devices remains challenging. Fabrication processes are often complex and prone to stochastic effects, such as unwanted pinning and thermal-induced instabilities, which limit device reliability and scalability. Addressing these challenges is crucial for advancing spintronic neuromorphic architectures toward real-world applications. Thus, to reduce these effects we have proposed a design which is experimentally feasible and require less energy as compared to existing one. By engineering the system anisotropy into a sawtooth-type energy landscape, we have achieved free flow of these microstructures and successfully emulated integrate and fire (IF) function of biological neuron. Thus, proposed design presents an experimentally reliable and energy efficient external stimuli approach for tailoring magnetic microstructures dynamic behaviours, resulting in low energy consumption of 23.66 $fJ$ per spike paving the way for the development of skyrmion-based futuristic neuromorphic computing device applications.

1. **Introduction:**

Magnetic microstructures such as Domain walls (DWs) and topologically protected skyrmions have emerged as potential candidates for spintronic-based applications, including logic gates, memory, and brain-inspired neuromorphic computing[1–3]. They have attracted a significant amount of attention due to their small dimensions, low energy requirement, along with easiness of their manipulation. Recent experiments and theoretical methods have shown that skyrmions can be stabilised in ferromagnetic (FM) multilayer films such as CoFeB/MgO, Co/Pt, and related stacks with the interplay of several magnetic interactions like interfacial perpendicular magnetic anisotropy (iPMA), interfacial Dzyaloshinskii-Moriya interaction (iDMI), and magnetic dipolar interactions[4–6].

To achieve various applications using these magnetic skyrmions, global researchers are putting huge efforts into using both experimental and theoretical methods for their creation[7], detection[8], protection during motion[9], and manipulation using different techniques[10,11]. Among various methods for direct observations of skyrmions, magneto-optical Kerr microscopy (MOKE) and magnetic force microscopy (MFM) remain the best and most convenient methods. However, from a practical applications point of view, these microstructures are detected using electrical methods, which produce electrical signals when these microstructures enter detection region based on magnetoresistance phenomena.

When manipulated by electrical currents, skyrmions face several challenges that hinder their practical implementation. Key drawbacks include Joule heating, which can cause device instability, and the skyrmion Hall effect (SkHE) arising from the Magnus force, which induces transverse motion and can lead to annihilation at device edges[12]. To reduce these effects, various strategies have been proposed, such as the voltage-controlled magnetic anisotropy (VCMA) technique[13,14], which leverages electric fields to modify anisotropy, as well as approaches based on mechanical strain[15] and engineered gradients[16,17]. These approaches are also well utilised for applications such as logic gates, sensors, memory devices, and brain inspired neuro computing. Out of these, brain inspired neurocomputing is one of the most researched fields as in which nanotrack with skyrmions are utilised as synapses as it offers low power consumption for data processing between pre synaptic neuron and post-synaptic neuron mimicking brain functions.

Numerous studies have investigated neuromorphic computing implementations using these microstructures, including skyrmion, skyrmionium and DWs as information carriers.

Techniques such as magnetic field modulation[11], Ruderman-Kittel-Kasuya-Yosida (RKKY) coupling modulation[18], PMA gradients[19,20], and DMI gradients[21] have been employed to enhance skyrmion dynamics. Current pulses (both Gaussian and square) have been used to optimize the behaviour of skyrmion-based architectures[22]. Additionally, artificial pinning sites have been introduced into nanotracks via local anisotropy modification, enabling controlled trapping and release of skyrmions. For DW-based systems, advanced fabrication techniques such as focused ion beam (FIB) irradiation and lithography have been used to tailor both structural and magnetic properties to meet specific device requirements[23]. Many of these studies have been carried out in material systems such as CoFeB/MgO[6,24], Co/Pd[25,26], and Co/Pt[27,28], owing to their strong PMA, tunable DMI, and experimental compatibility. Despite these advances, further work is needed to achieve precise, energy-efficient control of magnetic microstructures, particularly in simulation-driven exploration of neuromorphic concepts.

In this study, we implement gradient anisotropy engineering in nanotracks to enable energy-efficient motion of magnetic microstructures, allowing for lower power consumption and, in certain cases, spontaneous propagation without outside forces or polarized current pulses. The designed anisotropic landscape offers two functions: pinning locations for controlled trapping and inclined paths for free-flow motion. Depinning is accomplished by utilizing current pulses with well-calibrated temporal widths. This technique is investigated utilizing the Object Oriented MicroMagnetic Framework (OOMMF) on a Co/Pt nanotrack, where the anisotropy can be modified in regular 5 nm steps. Modern FIB techniques, such as $Ga^+$ irradiation, can precisely tune anisotropy at nanoscale scales, making it possible to achieve such nanometre resolution. This makes our proposed method not only theoretically viable but also experimentally feasible.

2. **Model and simulation details**

To determine the stable configuration of a magnetic material under an external field, micromagnetic simulation tools solve the Landau-Lifshitz-Gilbert (LLG) equation. Mathematically, the equation can be expressed as follows,

$$\frac{d\boldsymbol{m}}{dt} = -|\gamma_o|\boldsymbol{m} \times \boldsymbol{H}_{eff} + \alpha\left(\boldsymbol{m} \times \frac{d\boldsymbol{m}}{dt}\right) + \boldsymbol{T}_{CIP} \quad (1)$$

where $\boldsymbol{m}$ is the normalized magnetization, $\boldsymbol{m} = \boldsymbol{M}/M_s$. The effective magnetic field of the system under consideration is given by: $\boldsymbol{H}_{eff} = -1/(\mu_o M_s)(\partial E_{total}/\partial \boldsymbol{m})$ where $\mu_o$ is the permeability in vacuum. The third term in the above equation gives the spin transfer torque (STT) due to the spin-polarized current in the nanotrack. $H_{total}$ is the total magnetic energy of the system, which consists of contributing energy terms. The total Hamiltonian of the system can be given by,

$$H_{total} = H_{Ani} + H_{Ex} + H_d + H_z + H_{DMI} \qquad (2)$$

where, $H_{Ani}$ is the magneto-crystalline anisotropy energy, $H_{Ex}$ indicates the interaction of spins with the next neighbouring spin, $H_d$ indicates the dipolar nature of the individual magnetic particles that produce the stray field, and $H_z$ indicates an applied external field. We have performed micromagnetic simulations that also incorporate DMI, utilizing the finite-element difference (FED) approach. We used a Co film with a thickness of 0.4 nm on top of the Pt heavy metal to run these simulations. Nanotrack's geometrical dimensions were 400 × 40 × 0.4 nm, while the simulation's cell size was kept at 1 × 1 × 0.4 nm. The non-adiabaticity factor, $\beta = 0.3$, and the drift velocity of conduction electrons, $u = jPg\mu_B/(2eM_s)$, where $j$ is current density (in A/m$^2$), $P$ is the spin polarization equal to 0.4[29] for Co, $e$ is the electron charge, Landé g-factor $g$, and Bohr magnetron $\mu_B$, were taken into consideration as simulation parameters for simulating skyrmion dynamics.

In order to examine the skyrmion dynamics induced by current-in-plane (CIP) injection along the nanotrack's x-axis, the associated spin transfer torque $\boldsymbol{T}_{CIP}$ is written as follows:

$$\boldsymbol{T}_{CIP} = u\boldsymbol{m} \times \left(\boldsymbol{m} \times \frac{\partial \boldsymbol{m}}{\partial x}\right) + \beta u \boldsymbol{m} \times \frac{\partial \boldsymbol{m}}{\partial x} \qquad (3)$$

where the initial term describes the interaction between magnetic moments and the spin-polarized current $j$ through STT, whereas the other term captures the influence of non-adiabatic effects, governed by the non-adiabaticity factor ($\beta$). Below given table 1 below lists other simulation parameters that were considered during the simulations.

**Table I.** Magnetic Parameters Used in OOMMF Simulations

| Sr. No. | Parameters | Value | Units | Description |
|---|---|---|---|---|
| 1. | $A$ | $15 \times 10^{-12}$ | (J/m) | Exchange constant |
| 2. | $M_s$ | $5.8 \times 10^5$ | (A/m) | Magnetization |
| 3. | $\alpha$ | 0.3 | - | Gilbert damping parameter [30] |
| 4. | $D$ | 2.0 - 4.0 | (mJ/m$^2$) | DMI factor |
| 5. | $K_u$ | $0.08 – 1.2 \times 10^6$ | (J/m$^3$) | Anisotropy constant |

3. **Results and discussion:**

*3.1 Skyrmion Stability in Nanotracks Under Motion:*

The static and dynamic characteristics of skyrmions are strongly influenced by the geometry, dimensions, anisotropy, and DMI strength of the nanotrack[31]. Therefore, we began by identifying the optimal range of $K_u$ and DMI strength that ensures skyrmion stability throughout its motion across the entire nanotrack in our system. For skyrmion, a bubble domain state was defined, which transforms into typical Néel skyrmions. The bubble domain exhibits two distinct spin configurations, where the red and blue regions correspond to spins aligned in the +z and -z directions, respectively. In contrast, the white region in the centre represents spins oriented in the xy-plane, which delineates the domain wall width. Initially, the bubble domain possesses an inner and outer radius of 6 and 9 nm, respectively. Upon stabilization, this configuration evolves into a skyrmion structure. In this study, the effective size of the skyrmion ($r$) is defined based on the $m_z$ component, where $m_z = 0$. This approach ensures a consistent and precise measurement of the skyrmion dimensions[32].

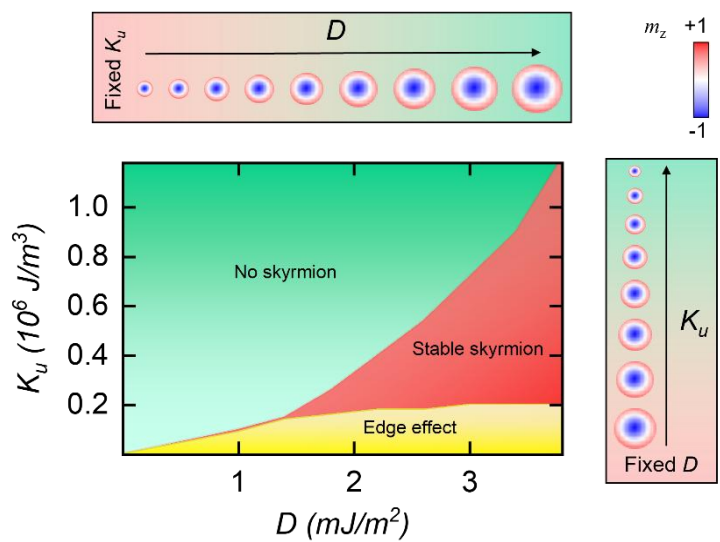

**Figure 1:** Phase diagram illustrating skyrmion stability under motion, mapped as a function of the $K_u$ (in J/m$^3$) and the $D$ (in mJ/m$^3$) values. The diagram delineates three distinct regions: stable skyrmion existence (red), absence of skyrmions (green), and skyrmion annihilation caused by edge effects (yellow). Insets provide additional insights: (Left) skyrmion radius ($r$) as a function of $K_u$ for a fixed $D$, (Top) skyrmion radius ($r$) as a function of $D$ for a fixed $K_u$.

To explore the relationship for stability and dynamics of skyrmion under CIP injection, we constructed a phase diagram shown in Figure 1, which maps regions of skyrmion stability within a nanotrack during its motion from one end to the other. The results reveal three distinct stability regimes. The red region denotes the region where skyrmion remain stable during motion, achieving an optimal balance between $D$ and $K_u$ that prevents collapse or any annihilation at edges. The particular combination of parameters was selected since it is easily adjustable with the help of external factors like strain or electric fields. The green region reflects conditions where skyrmion fail to form due to insufficient $D$ or excessively high $K_u$, suppressing nucleation. As DMI strength increases, skyrmion size grows (top inset, Fig. 1), while higher anisotropy reduces it (left inset, Fig. 1). The red region, therefore, highlights optimal conditions for stable skyrmion motion. The yellow region represents destabilization caused by edge effects, often due to larger skyrmion size at high $D$, making them vulnerable to edges. For $D \leq 1.2$, skyrmion fail to stabilize regardless of $K_u$, leading to saturation state of the nanotrack. At $D=1.4$, skyrmion exhibit transient stability for $K_u=0.14 \times 10^6$ J/m$^3$, however, they are eventually destroyed at the nanotrack edges, a process termed "edge annihilation," preventing them from completing their trajectory. It has been reported that, to prevent skyrmion from drifting from the direction of electron flow due to the Magnus force, we can use high anisotropy materials like SmCo$_5$ and Nd$_2$Fe$_{14}$B at the edges to confine the skyrmion in the nanotrack[33]. Thus, in our further results, we have only considered the results where the skyrmion completes its trajectory without any high $K_u$ material at the edges. However, for $D=1.4$, the skyrmion successfully completes its motion over the nanotrack for $K_u$ 0.15 × 10$^6$ J/m$^3$ without being annihilated at the margins. When $K_u$ is increased to 0.16 × 10$^6$ J/m$^3$, the skyrmion does not stabilize and gets saturated inside the nanotrack, suggesting that a very little change in anisotropy can completely change the existence and behaviour of the skyrmion.

*3.2 Effect of inclined energy landscapes on skyrmion dynamics:*

To understand how an inclined energy landscape influences skyrmion motion, the Thiele formalism can be used, where the skyrmion is considered as a rigid and incompressible particle. According to the Thiele equation[34]:

$$G \times v - \alpha D_F v + F_{ext} = 0 \qquad (4)$$

where, first term is called the gyroscopic term, which acts like the Magnus force caused by the transverse motion of the skyrmion during its motion, also called SkHE with $G = 4\pi Q \mu_o M_s t_{FM}/\gamma \boldsymbol{n_z}$ and $v$ is the velocity of the skyrmion. Here, $Q$ is the topological charge of the particle, which is $\pm 1$ in the case of a skyrmion. The second term in equation (4) is because of the damping term, which leads to the dissipation of energy, with $\alpha$ being gilbert damping, and $D_F$ is given by:

$$D_F = \begin{pmatrix} d_F & 0 \\ 0 & d_F \end{pmatrix}, d_F = \int dxdy \partial_i \boldsymbol{m} \, \partial_i \boldsymbol{m} \, ; x = y = i \qquad (5)$$

while the third term is the external force which drives the skyrmion into motion, which in our case is served by the PMA gradient, given by:

$$F_{ext} = u_F t_{FM} \frac{dK_u}{dx} \boldsymbol{n_x} \qquad (6)$$

where $u_F = \int dxdy(1 - m_z^2)$. Therefore, considering $Q = 1$ for skyrmion and using equations (5) and (6), the velocity of skyrmion due to gradient anisotropy is given by:

$$v = \sqrt{v_x^2 + v_y^2} \tag{7}$$

$$v_x = \frac{\alpha d_F \gamma u_F}{\mu_o M_s [16\pi^2 + (\alpha d_F)^2]} \frac{dK_u}{dx} \tag{8}$$

$$v_y = \frac{4\pi \gamma u_F}{\mu_o M_s [16\pi^2 + (\alpha d_F)^2]} \frac{dK_u}{dx} \tag{9}$$

When a skyrmion moves within a confined nanotrack, the edge effects become significant and can contribute a nonzero potential acting on the skyrmion. In our simulations, we have considered linear motion only along the direction of the anisotropy gradient. Thus, this motion is balanced by the drive force and the edge-induced potential. Under this condition, the velocity component of the skyrmion along the y-axis can be considered negligible, i.e., $v_y$=0. Therefore, the velocity of the skyrmion can be given by[35]:

$$v = v_x = \frac{\gamma u_F}{\mu_o M_s \alpha d_F} \frac{dK_u}{dx} \tag{10}$$

Here, both $d_F$ and $u_F$ are dependent on the skyrmion profile, and they can be assumed to be:

$$u_F \approx 4\pi r C, d_F \approx 2\pi (r/C + C/r) \tag{11}$$

where $C$ is given by $\approx \Delta/\sqrt{\Delta^2/r^2 + 1}$ and $r$ is skyrmion radius, which is given by:

$$r \approx \frac{\Delta}{\sqrt{2 - 2 D/D_c}} ; D_c = 4\sqrt{AK}/\pi \tag{12}$$

The skyrmion velocity can be expressed in a simplified analytical form as follows, by replacing the equations for $u_F$ and $d_F$ from Equation (11), as well as the skyrmion radius $r$ from Equation (12), in the velocity expression:

$$v_x = \frac{2\gamma A}{\mu_o M_s \alpha K \left(5 - 4 D/D_c\right)} \frac{dK_u}{dx} \tag{13}$$

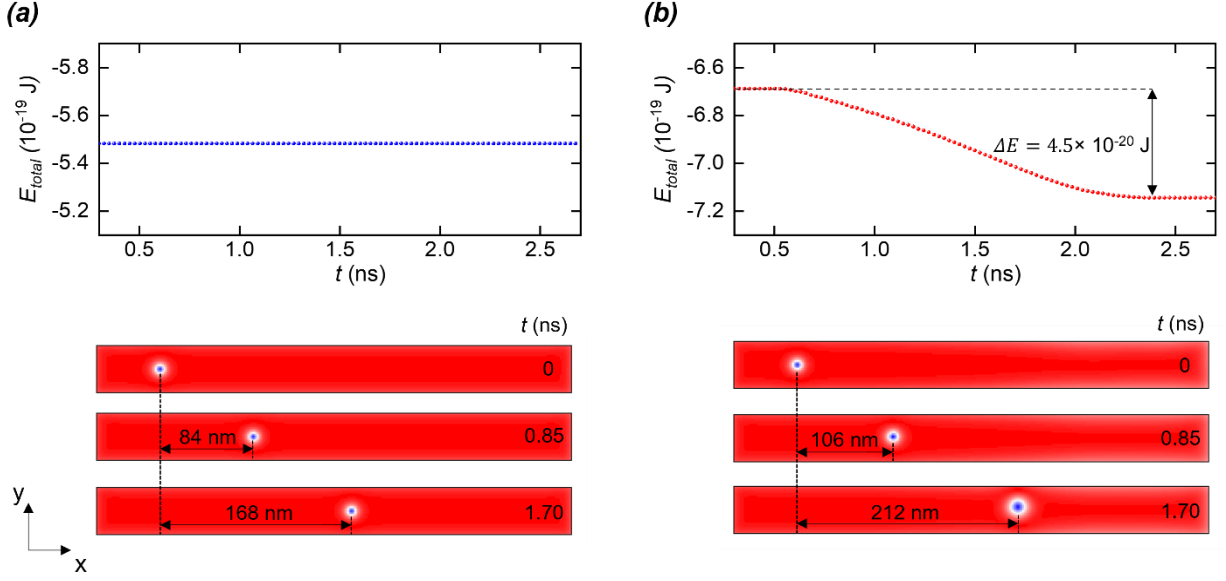

**Figure 2:** (a) Total energy of the system during the motion of a skyrmion using $j = 2.5 \times 10^{12}$ A/m$^2$ along the nanotrack with constant $K_u=0.7 \times 10^6$ J/m$^3$ and $D=3$. (b) Total energy of the system when the anisotropy of the nanotrack is gradually varied from $K_u=0.7$ to $K_u=0.3 \times 10^6$ J/m$^3$ with $D=3$. The snapshots below both plots show the skyrmion positions in the nanotrack at 0, 0.85, and 1.70 ns, highlighting its displacement under the respective conditions.

Therefore, after performing stability simulations for skyrmions in a Co/Pt nanotrack, we extended our study to explore their dynamic behaviour during motion. Initially, we investigated the skyrmion velocity in a nanotrack with a uniform DMI strength of 3 mJ/m$^2$ and constant $K_u = 0.7 \times 10^6$ J/m$^3$ under a current density of $j = 2.5 \times 10^{12}$ A/m$^2$. Under these conditions, the skyrmion achieved a velocity of approximately 97.14 m/s. To examine the energetic landscape during motion, we analyzed the total energy of the system as defined in Equation (2). Since both $D$ and $K_u$ were uniform throughout the nanotrack, the skyrmion maintained a stable shape and trajectory, as reflected in Figure 2(a). The total system energy remained constant at approximately $-5.48 \times 10^{-19}$ J throughout the motion of the skyrmion.

Inspired by physical systems' tendency to choose lower-energy states, we added a spatial variation in the $K_u$ around the nanotrack length. The goal of this strategy was to introduce an energy gradient in order to influence and control skyrmion motion. Based on the calculations performed to evaluate skyrmion stability while in motion, the size of the skyrmion never reduced below 9.5 nm. This led to the selection of a 5 nm step size for the anisotropy variation in these calculations. This resolution is practically feasible using modern fabrication techniques in multilayer thin films, particularly FIB irradiation, which enables precise modification of magnetic properties within regions as small as 5 nm by controlling beam focus and intensity. With the introduction of $K_u$ variations along the nanotrack, noticeable changes in both energy and skyrmion dynamics were observed. As shown in Figure 2(b), which shows the total energy as a function of time, the system energy decreases during skyrmion motion, reaching a net reduction of $\Delta E = -4.5 \times 10^{-20}$ J. This energy gradient has significantly influenced the skyrmion behaviour. Under uniform $K_u$, the skyrmion travels 84 nm in 0.85 ns while preserving its original size. In contrast, when $K_u$ is spatially varied, the skyrmion covers 106 nm within the same time, corresponding to an increase in velocity from 97.14 m/s to 120.33 m/s. Additionally, the skyrmion slightly enlarges in regions with reduced anisotropy, consistent with its known dependence on the local energy landscape.

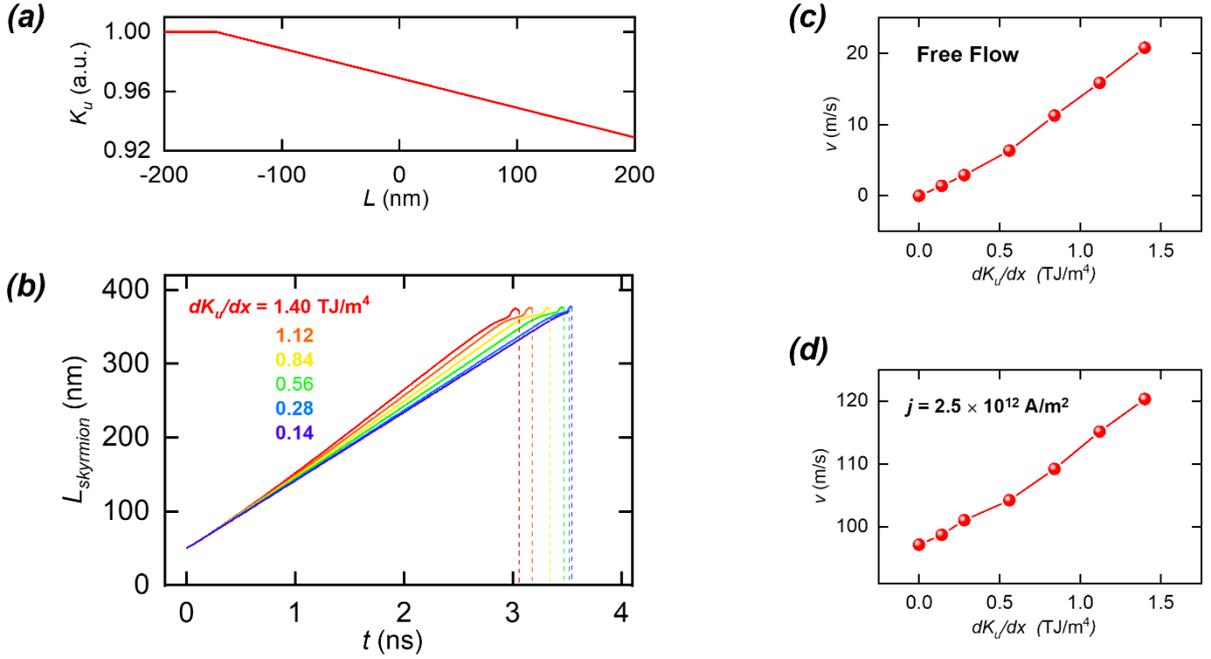

**Figure 3: (a)** Variation of $K_u$ across the nanotrack. The maximum value is $K_{u-max} = 0.7 \times 10^6$ J/m², while the minimum $K_{u-min}$ is varied to give different gradients ($dK_u/dx$) within the range of 50 nm to 400 nm, illustrating the tunable energy landscape. **(b)** Position and trajectory of a skyrmion under current density $j = 2.5 \times 10^{12}$ A/m², indicating directed motion along the track and the time ($t$ in ns) required to complete the motion. **(c)** Skyrmion motion in the absence of external drive, showcasing free-flow dynamics influenced entirely by the energy landscape gradients. **(d)** Enhanced skyrmion velocity arising from the combined effects of applied current ($j = 2.5 \times 10^{12}$ A/m²) and the energy gradient, demonstrating an additive contribution to net motion.

To further understand the influence of a gradient energy landscape on skyrmion dynamics, we examined various anisotropy gradients with maximum value is $K_{u-max} = 0.7 \times 10^6$ J/m², while the minimum $K_{u-min}$ is varied to give different gradients ($dK_u/dx$) within the range of 50 nm to 400 nm, illustrating the tunable energy landscape in Figure 3 (a). It was observed that for current density $j = 2.5 \times 10^{12}$ A/m², under different gradients, from 0.14-1.40 TJ/m⁴, the skyrmion was able to complete its motion in shorter transit times for the skyrmion, indicating enhanced motion with increasing gradient strength, as shown in Figure 3(b). To isolate the effect of the energy gradient alone, additional simulations were carried out with zero current density. Remarkably, even in the absence of external driving forces, the skyrmion exhibited spontaneous motion freely flowing toward regions of lower anisotropy. This confirms that the anisotropy gradient itself acts as a driving mechanism. At the highest gradient value of 1.40 TJ/m⁴, the skyrmion achieved a velocity of up to 20 m/s, as shown in Figure 3(c). Hence, these findings demonstrate that the engineered anisotropy gradient acts as an effective driving force, enhancing skyrmion velocity. The combined effect of the applied current and the gradient-induced force is summarized in Figure 3(d), which shows the skyrmion velocity resulting from applied current density and energy landscape-induced motion.

*3.3 Possible application using skyrmion-based neuron system*

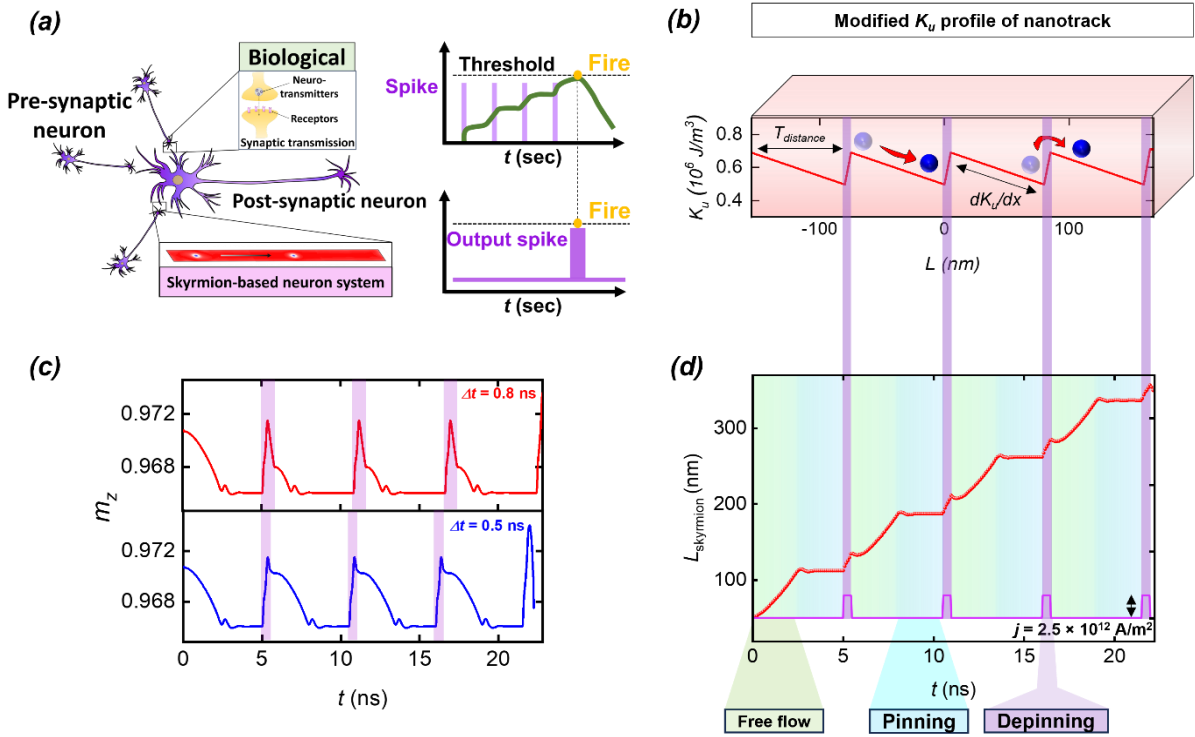

**Figure 4: (a)** Schematic illustration of a biological neuron, showing synaptic transmission between pre- and post-synaptic neurons via neurotransmitters. An analogous mechanism is mimicked in a skyrmion-based nanotrack, where skyrmion motion plays a similar role in information transfer. **(b)** The diagram displays the modified $K_u$ profile in a 400 nm long nanotrack. Here, the blue sphere represents a skyrmion, which is pushed upward by an external force across a $K_u$ step, then, due to an anisotropy gradient of 3.0 TJ/m⁴ ($dK_u/dx$ in TJ/m⁴), the skyrmion is moved freely along the nanotrack. Distance between consecutive tooth ($T_{distance}$) is 70 nm. **(c)** Magnetization state ($m_z$) versus time ($t$ in ns) plot corresponding to different pulse widths of $j = 2.5 \times 10^{12}$ A/m² **(d)** Plot shows skyrmion location ($L_{Skyrmion}$) versus simulation time (in ns) along with pulse current density $j = 2.5 \times 10^{12}$ A/m² used for 0.5 ns (shown in bottom inset in purple curve).

Figure 4 (a) shows a schematic of the biological neuron system, where between the pre-synaptic neuron and post-synaptic neuron, information is transferred using neuro-transmitters and received by receptors. Same biological functionality can be mimicked by a skyrmion-based neuro system, where skyrmions work as neuro-transmitters between the pre- and post-neurons. Perfect integrate and fire neuron is one of the most analysed behaviours of the nervous system by global researchers by both experimentally and theoretically, using simulation tools, which treats the neuron as a circuit that integrates input spikes until a threshold membrane potential is reached, after which it fires an output spike and resets its voltage. This time-dependent spike function ($S(t)$) of the skyrmion-based neuro system is given by:

$$S(t) = f\left(\sum w_i(t) j_i(t) + k_o\right) \qquad (14)$$

where $k_o$ indicates the initial position (i.e., membrane potential) of the synaptic skyrmion, $w_i(t)$ stands for time synaptic weight, and $j_i(t)$ for the input spike current density of the ith synapse.

As discussed in the previous section, a step size of 5 nm was chosen for anisotropy variation, which is smaller than the typical skyrmion diameter under all conditions. The proposed skyrmion-based neuron computing device, shown in Figure 4(b), uses a sawtooth-shaped anisotropy $K_u$ profile along the nanotrack. Such spatially periodic anisotropy engineering is experimentally feasible via the FIB irradiation technique, wherein controlled ion fluence enables precise modulation of magnetic anisotropy at nanometric scales. This methodology not only facilitates deterministic tailoring of the $K_u$ profile but also intrinsically suppresses defect formation, a limitation commonly associated with conventional lithographic and etching-based patterning processes. The introduction of a sawtooth anisotropy landscape serves to engineer the system's internal energy contributions, thereby establishing an inclined potential profile optimized for low-energy operation in spintronic-based brain-inspired applications. Among different approaches used by researchers, anisotropy modulation represents a robust and cost-effective method for introducing inclined energy landscapes within FM nanostructures. Within this proposed design, a skyrmion is nucleated at the left side of the track, specifically at 50 nm, corresponding to the initial position $k_o$ as defined in equation 14. The skyrmion first moves freely without any external drive until stabilized at the subsequent sawtooth potential minimum. Overcoming this metastable pinning state necessitates the application of an external drive, thus, a current pulse of appropriate width and amplitude is applied to push the skyrmion over the barrier.

To start with, a current density of $j = 2.5 \times 10^{12}$ A/m² with different pulse widths was applied to examine whether the skyrmion could overcome the anisotropy barrier with $K_{u\text{-}max}$ and $K_{u\text{-}min}$ values of 0.70 and 0.49 × 10⁶ J/m², respectively. Figure 4(c) presents the variation of $m_z$ state as a function of time ($t$, in ns) for the nanotrack, showing changes in $m_z$ that arise from variations in skyrmion size. It is well established, and further confirmed by these results, that the skyrmion expands in size when moving from a region of higher anisotropy to a region of lower anisotropy. Consequently, the $m_z$ state also changes. Initially, the skyrmion was allowed to move freely until reaching a low-energy region, where it stabilized at the sawtooth potential minimum. As the current density pulse was applied, the skyrmion was suddenly pushed into the higher anisotropy region from the low-anisotropy region. The suddenness of the current pulse triggered this immediate transition, which resulted in an instant change in skyrmion size. This change was reflected in the $m_z$ state profile as a peak-like feature. From Figure 4(c), it can be observed that the duration of the current pulse plays a critical role in determining the skyrmion dynamics. When a relatively long pulse of 0.8 ns was applied, the skyrmion was not only pushed into the higher anisotropy region but also driven across the inclined anisotropy slope. This resulted in nearly symmetric changes in the $m_z$ state. In contrast, when a shorter pulse of 0.5 ns was applied and then switched off, the skyrmion was displaced into the higher anisotropy region but subsequently allowed to move freely along the inclined slope. This produced asymmetric variations in the $m_z$ state. These results indicate that the motion of skyrmions across sudden anisotropy barriers induces sharp and controllable changes in skyrmion size. Such abrupt modulations can be harnessed in future spintronic devices, particularly for applications requiring fast, energy-efficient signal generation, non-linear magnetization responses, and neuron-inspired computing functionalities, especially in systems where spike-like features are essential, as reflected in the $m_z$ state changes observed here.

Figure 4(d) illustrates the motion of a skyrmion under a 0.5 ns current pulse, which shows three different dynamical regions, free flow, pinning or stabilization, and subsequent depinning caused by the applied current pulse. Neuronal dynamics can be directly matched to this behaviour, where the free flow represents integration of input signals, the pinning corresponds to temporary storage or accumulation of potential, and the depinning mimics the continuous skyrmion motion, until it reaches the detector region, which corresponds to the threshold condition in biological neurons. To illustrate an integrate-and-fire neuromorphic function, four successive sawtooth anisotropy profiles were used in the present design. Every sawtooth step serves as a synaptic weight stage, and the mechanism initiates a firing event and resets the skyrmion position when the skyrmion reaches the detector region. Although this study used four stages, the design can easily be expanded to allow for advanced neuron-inspired architectures by including a greater number of stages based on the required stages. In addition to this, we have also evaluated the energy consumption associated with information transmission in the proposed skyrmion-based neuron system. The energy required to drive a skyrmion over a single stage is estimated using:

$$E = \rho\, t_{HM}\, l\, w\, j^2 T_{\text{pulse}} \qquad (15)$$

where $t_{HM}$ is the thickness of the HM layer, $l$ and $w$ are the length and width of the nanotrack, $j$ is the applied pulse density, and $T_{pulse}$ is the driving pulse duration. As per literature[36], for Pt layer of thickness 5 nm, the resistivity is nearly $\rho = 2 \times 10^{-7}$ Ωm. Therefore, using $l = 400$ nm, $w = 40$ nm, $j = 1.72 \times 10^{12}$ A/m², and $T_{pulse} = 0.5$ nsec, the energy consumption per spike is 23.66 fJ/spike. This low energy cost per spike highlights the suitability of such graded skyrmion tracks for neuromorphic and in-memory computing systems, where energy efficiency is crucial. Further, based on these sawtooth steps, we have also extracted the minimum current density amplitude required to depin the skyrmion and continue its motion to the next inclined slope. Table II summarizes the values of $K_{u\text{-}max}$ and $K_{u\text{-}min}$, along with the corresponding gradient values, together with the required depinning current density ($j_{depin}$).

**Table II:** Summary of maximum $K_{u\text{-}max}$ and minimum $K_{u\text{-}min}$ (in 10⁶ J/m²) values used for simulation. The corresponding anisotropy gradient (in TJ/m⁴) and the depinning current density $j_{\text{depin}}$ (in 10¹² A/m²) required for a 0.5 ns pulse width are also listed.

| Sr. No. | $K_{u\text{-}max}$ (× 10⁶ J/m²) | $K_{u\text{-}min}$ (× 10⁶ J/m²) | $dK_u/dx$ (TJ/m⁴) | $j_{\text{depin}}$ (× 10¹² A/m²) |
|---|---|---|---|---|
| 1. | 0.70 | 0.59 | 1.50 | 1.72 |
| 2. | 0.70 | 0.56 | 2.00 | 1.97 |
| 3. | 0.70 | 0.52 | 2.50 | 2.25 |
| 4. | 0.70 | 0.49 | 3.00 | 2.47 |

*3.4 Possible application using DWs motion for perfect integrate and fire function:*

For showing DWs based on an artificial skyrmion neuron, we have used the same $K_u$ profile (shown in the top plot of Figure 5 (a)), which was used earlier for showing dynamics of skyrmion, i.e., $K_{u\text{-}max} = 0.70 \times 10^6$ J/m², and $K_{u\text{-}min} = 0.49 \times 10^6$ J/m², which gives $dK_u/dx$ value of 3.0 TJ/m⁴. In DW's cases as well, these sawtooth act as artificial pinning points. As mentioned earlier, the main advantage of this proposed device is that it does not require any conventional lithography or etching techniques to introduce notches or pinning sites, which often lead to unwanted defects. And hence DWs can be trapped by introducing artificial potential wells created by these sawtooth. Figure 5 (a) shows the $m_z$ states of DWs during their motion along the engineered nanotrack with time, where it demonstrates the integration and firing function of neurons. The zoomed-in image of Figure 5 (a) is shown below in Figure 5 (b), highlighting the three different regions during DW's motion. At (i) $t=0$ ns, DW is set to move freely towards the low energy region with a velocity of approximately 10 m/s and reaches point (ii) at $t=5.93$ ns. Due to the high $K_u$ value caused due to sawtooth, DW is pinned at this site till the depinning current pulse is applied. At point (iii) shown in Figure 5 (b), a current pulse of amplitude $2.5 \times 10^{12}$ A/m² is applied for 0.5 ns, so that it can depin DW with a velocity of approximately 70 m/s. Again, from $t$=9.57 to 13.87 ns, DW is moved freely under the effect of gradient $K_u$, and struck at the same $m_z$ state till the next pulse is applied. The transition of magnetization in these states is shown in Figure 5 (c) as a function of time. Figure 5(d) shows the correlation between the $dK_u/dx$ and the $j_{depin}$, defining two different areas, i.e., DW pinning and depinning, which have been caused by the artificial sawtooth energy barriers. It can be observed that as the $dK_u/dx$ is increased, it requires a higher current density. Interestingly, before reaching the pinning site, DWs can propagate under the influence of current densities below $j_{depin}$ with much higher velocity than free flow. Thus, highlighting a method for accelerating DW motion before controlled depinning.

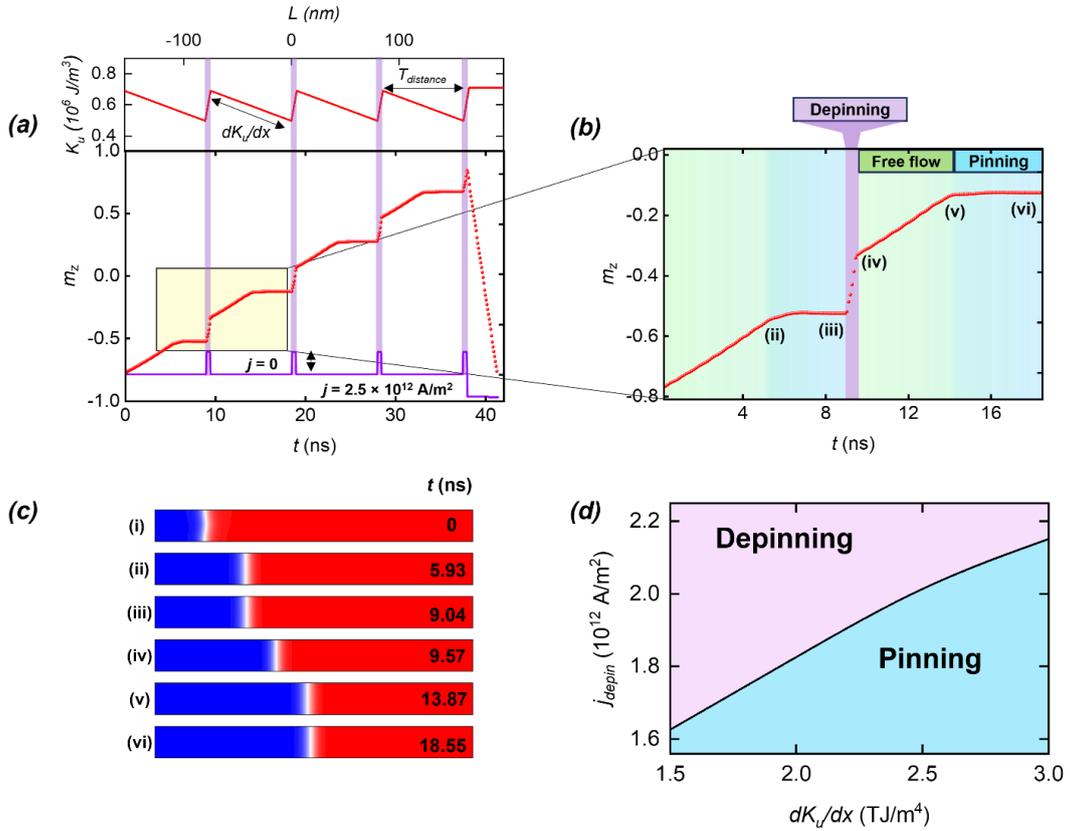

**Figure 5:** (a) Illustration of sawtooth-like $K_u$ profile used for DWs simulations with maximum value of $K_{u\text{-}max}$= 0.70 × 10⁶ J/m², while the minimum $K_{u\text{-}min}$ is = 0.49 × 10⁶ J/m², to give gradients (d$K_u$/dx) of 3.0 TJ/m⁴. Distance between consecutive tooth ($T_{distance}$) is 70 nm. Below the $K_u$ profile, the plot shows $m_z$ with simulation time (in ns) along with pulse current density $j$ = 2.5 × 10¹² A/m² used for 0.5 ns (shown in bottom inset in purple curve). **(b)** Zoom-in plot for the $m_z$ vs $t$ from 2.2 to 18.6 ns, showing depinning of DW (highlighted in purple), free flow (highlighted in green), pinning effect due to sawtooth (highlighted in sky-blue), and depinning of DWs **(c)** DWs snapshots during its motion at different intervals of time from $t$ = 0 - 18.55 ns. **(d)** Phase diagram illustrating the pinning and depinning regions of DWs as a function of the $dK_u/dx$ and the depinning current density ($j_{depin}$). It highlights the minimum current pulse necessary to depin the DW from the engineered energy barrier.

## 4. Conclusions:

In conclusion, we demonstrated that FM microstructures such as skyrmions and DWs can be energy efficiently driven by inclined energy landscapes, which in our proposed work is served by anisotropy modifications. Here, we proposed sawtooth-like energy landscapes which serve two main functionalities, first is the free flow of skyrmions and DWs, which helped to drive these microstructures' energy efficiently, and second is the step of successive tooth, which was utilised to pin these microstructures at those sites until a current pulse of appropriate amplitude and width is applied. It was also observed that, with an increase in anisotropy-modified energy landscapes values, the depinning current pulse also increased. Based on our simulation, we have proposed four anisotropy-modified regions, which work as four different resistive states in our proposed neuron-based device. Thus, using this anisotropy-modified resistive states, it helps us achieve a perfect integrate and fire function for neuron devices. Thus, our simulation-based work unveils an energy-efficient and low-power-consuming method of manipulating the FM microstructures, which is easy to implement experimentally using advanced FIB techniques and is helpful in futuristic spintronic applications, including logic, memory devices, and brain-inspired neuron-based devices.

**Acknowledgments**

AS is a Junior Research Fellow (JRF) supported by UPES, Dehradun. The authors would like to thank UPES Dehradun for providing research related infrastructure. The authors acknowledge the financial support from UGC-DAE CSR through a Collaborative Research Scheme (CRS) project number CRS/2022-23/01/675 as well as UPES-SEED grant (UPES/R&D-SoAE/25062025/17).

**CRediT authorship contribution statement**

**Anmol Sharma:** Writing-original draft, Writing-review & editing, Formal analysis, Visualization, Software. **Ranjeet Kumar Brajpuriya:** Writing-review & editing, Visualization, Validation, Data curation. **Vivek K. Malik:** Writing-review & editing, Visualization, Validation. **Vishakha Kaushik:** Writing-review & editing, Visualization, Validation. **Sachin Pathak:** Writing-original draft, Writing-review & editing, Visualization, Validation, Software, Supervision, Methodology, Investigation, Formal analysis, Data curation, Conceptualization.

**Data availability**

Data will be made available on request.